\RequirePackage{color}  
\documentclass{PoS}
\usepackage{amsmath}
\usepackage{booktabs}
\usepackage{bm}
\graphicspath{{figs/}}

\newcommand{\Vcb}{|V_{cb}|}
\newcommand{\Vub}{|V_{ub}|}
\newcommand{\Vcs}{V_{cs}}
\newcommand{\Vcd}{V_{cd}}
\newcommand{\bra}[1]{\langle #1 |}
\newcommand{\ket}[1]{| #1 \rangle}
\definecolor{dkgreen}{rgb}{0.0,0.4,0.0}

\def\[{\left[}
\def\]{\right]}
\def\({\left(}
\def\){\right)}
\newcommand{\matrixel}[3]{\left< #1 \vphantom{#2#3} \right|
 #2 \left| #3 \vphantom{#1#2} \right>} 

\title{$B$-meson semileptonic decays with highly improved staggered quarks}

\ShortTitle{$B_{(s)}$ semileptonic decays with HISQ}

\author{\speaker{Andrew Lytle}$^{,1}$\\
    E-mail: \email{atlytle@illinois.edu}\\
}
    
\author{
Carleton DeTar$^2$,
Aida El-Khadra$^1$,
Elvira G{\'a}miz$^3$,
Steven Gottlieb$^4$,
William Jay$^5$,
Andreas Kronfeld$^6$,
James Simone $^6$,
and Alejandro Vaquero$^2$
\newline\newline
Fermilab Lattice and MILC Collaborations
\newline
\\
$^1$ Department of Physics and Illinois Center for Advanced Studies of the Universe, University of Illinois, Urbana, Illinois, 61801, USA\\
$^2$ Department of Physics and Astronomy, University of Utah, Salt Lake City, Utah 84112, USA\\
$^3$ CAFPE and Departamento de Física Teórica y del Cosmos, Universidad de Granada, E-18071
Granada, Spain\\
$^4$ Department of Physics, Indiana University, Bloomington, Indiana 47405, USA\\
$^5$Center for Theoretical Physics, Massachusetts Institute of Technology, Cambridge, MA 02139, USA\\
$^6$ Theory Division, Fermi National Accelerator Laboratory, Batavia, Illinois, 60510, USA\\
}

\abstract{
We present an update of the Fermilab Lattice and MILC Collaborations project to compute the
form factors for semileptonic $B_{(s)}$-meson decays.
Our calculation uses the highly improved staggered quark (HISQ) action for sea and valence quarks, and ensembles with up, down, strange, and charm quarks in the sea. 
Using a highly improved action with the MILC Collaboration's gauge ensembles with lattice spacings down to $a\approx0.03$ fm, allows the heavy valence quarks to be treated with the same discretization as the light and strange quarks.
This unified treatment of the valence quarks allows for absolutely normalized vector currents, bypassing the need for perturbative matching, which has been a source of uncertainty in previous calculations of $B$-meson decay form factors by our collaboration.
All preliminary form-factor results are blinded.\\

FERMILAB-CONF-23-017-T \\
MIT-CTP/5522
}

\FullConference{%
The 39th International Symposium on Lattice Field Theory,\\
8th-13th August, 2022,\\
Rheinische Friedrich-Wilhelms-Universität Bonn, Bonn, Germany}

\begin{document}
\section{Introduction}
Lattice QCD calculations of hadronic decay form factors are critical inputs for high precision tests of the Standard Model (SM) of particle physics. Ab-initio predictions for hadronic
flavor-changing matrix elements allow for the extraction of CKM matrix elements from experimentally measured decay rates, which in turn enable precision tests of the Standard Model. 

Recent years have witnessed a resurgence of theoretical interest in the heavy flavor sector, driven by new results from experiments at LHC, and anticipation of new results from Belle II.
 Several few-sigma discrepancies (collectively referred to as $B$-anomalies) may be hints for new physics, and ongoing experiments
continue to reduce experimental uncertainties, enabling sharper
tests of the SM, provided theory uncertainties are quantified
at a commensurate level.
There are also the long-standing differences in inclusive and exclusive extractions of the CKM elements $\Vub$ and $\Vcb$ which should be resolved. From the theory side, improved theoretical calculations of $B$-semileptonic decays will bear directly both on exclusive/inclusive discrepancies, and interpreting $B$-anomalies. 
(For recent discussions from a lattice perspective, see e.g.~\cite{Boyle:2022uba,USQCD:2022mmc,Davoudi:2022bnl,Vaquero:2022mak}.)
 In this proceeding, we provide a status update of the FNAL-MILC collaboration's calculations of semileptonic $B_{(s)}$-meson decay form factors using the highly improved staggered quark (HISQ) action.

\section{Calculation overview}
Our calculation uses ensembles generated by the MILC Collaboration using $N_f=2+1+1$ flavors of dynamical sea quarks with the HISQ action~\cite{MILC:2010pul,MILC:2012znn,Bazavov:2017lyh}.
Here we show results from ensembles with lattice spacings of $a \approx 0.09$, 0.06, and 0.042 fm. At $a \approx 0.09$ and 0.06 fm we have generated correlator data
on ensembles with light sea-quarks at their physical values as well as at $m_l/m_s = 0.1, 0.2$.
At $a\approx 0.042$ fm we have analyzed an ensemble with $m_l \approx 0.2 m_s$ in the sea, and we plan to include a physical-mass ensemble in the future.
The strange and charm sea-quark masses are tuned to be close to their physical values, and the valence light- and strange-quark masses are taken to be equal to the corresponding sea-quark masses.
The heavy valence quarks range in mass from roughly $0.9 m_c$ to just below the lattice cutoff $a m_h  \lesssim 1$.
At the finest lattice spacings of $a\approx 0.042$~fm and 0.03 fm, this setup allows simulation close to the physical mass of the bottom quark.

\begin{figure}[t]
    \centering
    \includegraphics[width=0.45\textwidth]{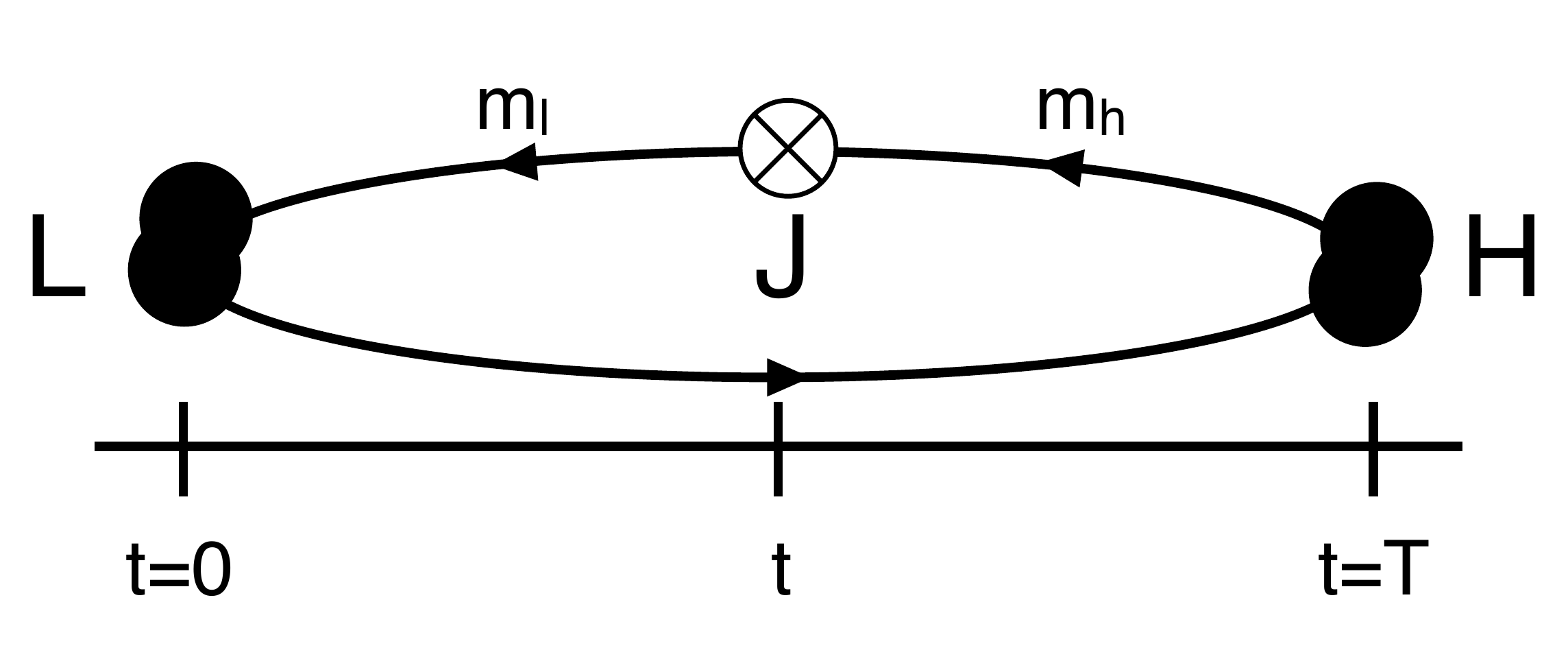}
    \caption{A schematic figure of the 3pt functions defined in Eq.~(\ref{eq:3pt}).
    The light final state hadron is created with momentum $\bm{p}$ at the origin.
    An external current $J$ is inserted at time $t$.
    The heavy initial hadron is destroyed at rest at time $T$.
    }
    \label{fig:schematic_3pt}
\end{figure}

To determine the form factors, we compute the following two-point and three-point correlation functions:
\begin{align}
    C_H(t)
    &= \sum_{\bm{x},\bm{y}}
    \left\langle
        \mathcal{O}_H(t_{\text{src}}, \bm{x})
        \mathcal{O}_H(t+t_{\text{src}}, \bm{y})
    \right\rangle \label{eq:initial_2pt}\\
    C_L(t,\bm{p})
    &= \sum_{\bm{x},\bm{y}}
    e^{i \bm{p}\cdot(\bm{x}-\bm{y})}
    \left\langle
        \mathcal{O}_L(t_{\text{src}}, \bm{x})
        \mathcal{O}_L(t+t_{\text{src}}, \bm{y})
    \right\rangle \label{eq:final_2pt}\\
    C_3(t,T,\bm{p})
    &= \sum_{\bm{x},\bm{y},\bm{z}}
    e^{i \bm{p}\cdot(\bm{x}-\bm{y})}
    \left\langle
        \mathcal{O}_L(t_{\text{src}}, \bm{x})
        J(t+t_{\text{src}}, \bm{y})
        \mathcal{O}_H(T+t_{\text{src}}, \bm{z})
    \right\rangle, \label{eq:3pt}
\end{align}
where the $\mathcal{O}_{H,L}$ are staggered meson operators which couple to the heavy initial ($H$) and light final state ($L$) hadrons.
For the scalar and temporal vector current we employ local staggered operators, while for spatial vector current we use the one-link operator.
For brevity, we have suppressed staggered structure and Lorentz indices in the lattice current $J$, which represents the scalar, vector, and tensor currents.
Figure~\ref{fig:schematic_3pt} shows the structure of these correlation functions.
We work in the rest frame of the decaying hadron $H$ and compute the recoiling hadron $L$ with eight different lattice momenta $\bm{p}_L = \frac{2\pi}{N_s a} \bm{n}$, where $N_s \in \mathbb{Z}$ is the spatial extent of the lattice and $\bm{n}$ is $(0,0,0)$, $(1,0,0)$, $(1,1,0)$, $(2,0,0)$, $(2,1,0)$, $(3,0,0)$, $(2,2,2)$ or $(4,0,0)$.
For each choice of mass and momentum, we compute the three-point functions for a few (typically 4 or 5) different source-sink separations $T$.
For the light-quark propagators in the calculation, we employ the truncated solver method~\cite{Bali:2009hu}, using 24 to 36 loose solves per configuration.

To extract the required matrix elements, our analysis employs joint correlated fits to the two-point and three-point correlation functions using the spectral decomposition.
For instance, for the three-point function in Eq.~(\ref{eq:3pt}), the spectral decomposition reads
\begin{align}
    C_3(t,T,\bm{p})
    = \sum_{m,n}
        (-1)^{m(t+1)} (-1)^{n(T-t-1)}
        A_{mn}
        e^{-E_L^{(n)}(\bm{p})t}
        e^{-M_H^{(m)}(T-t)}.\label{eq:3pt_spectral_decomp}
\end{align}
As usual for staggered fermions, the correlation functions include smoothly decaying contributions with the desired parity as well as oscillating contributions from states of opposite parity.
The spectral decompositions for the two-point functions are similar.
The ground-state amplitude $A_{00}$ is proportional to the matrix element $\matrixel{L}{J}{H}$, and so a fit to Eq.~(\ref{eq:3pt_spectral_decomp}) gives the required matrix element.
For the sake of visualization, the following ratio of correlation functions is useful:
\begin{align}
R^S(t, T, \bm{p}) =
    \sqrt{2 M_H} \left(\frac{m_h - m_l}{M_H^2 - M_L^2}\right)
    \frac{ C^S_3(t,T, \bm{p})}{ \sqrt{C_2^L(t, \bm{p}) C_2^H(T-t) e^{-E_L t} e^{-M_H (T-t)}}}.
\label{eq:ratio}
\end{align}
Up to discretization effects, this ratio asymptotically approaches the form factor $f_0$ for large times:
\begin{align}
    R^S(t, T, \bm{p}) \stackrel{0 \ll t \ll T}{\longrightarrow} f_0(\bm{p}).
\end{align}
Slightly different ratios, differing only by kinematic prefactors and renormalization factors, can also be constructed for $f_\parallel$ and $f_\perp$.
Although our quantitative analysis is based on fits to the spectral decomposition, the ratio provides a valuable visual check on the results.

All of our results are blinded by a random factor that is common for all three-point functions of a given analysis/decay channel.
We will carry the analysis of the blinded form factors all the way through the chiral interpolation and continuum extrapolation, unblinding only when the analysis of systematic errors is complete.

\section{Results}
\subsection{Two-point and three-point functions}
\begin{figure}[]
\centering
\includegraphics[width=0.8\textwidth]{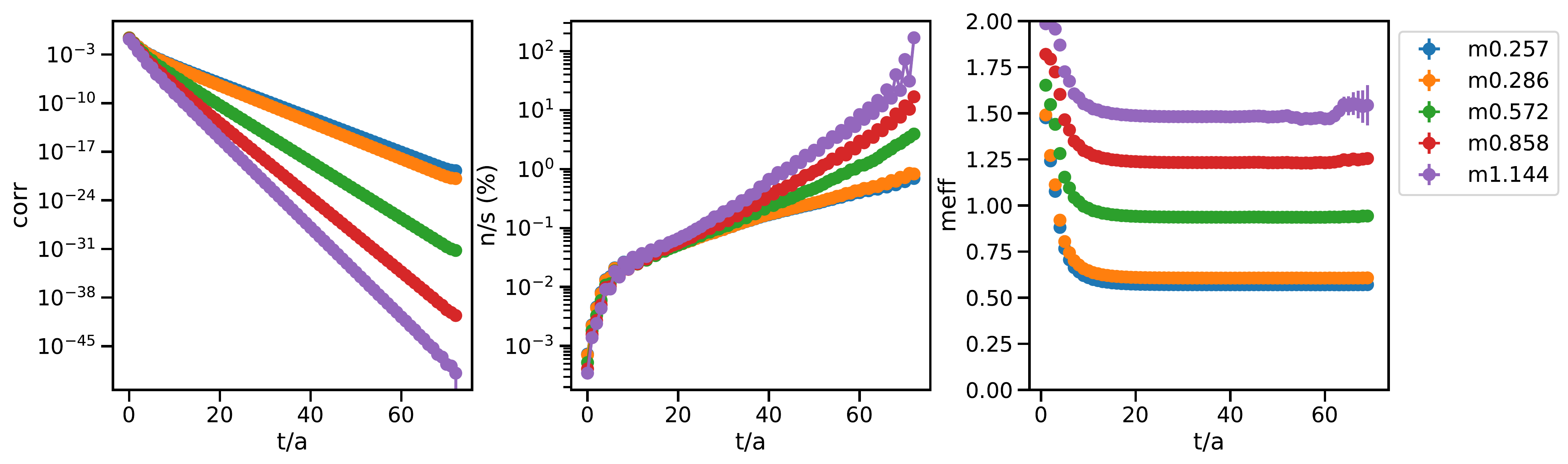}
\includegraphics[width=0.8\textwidth]{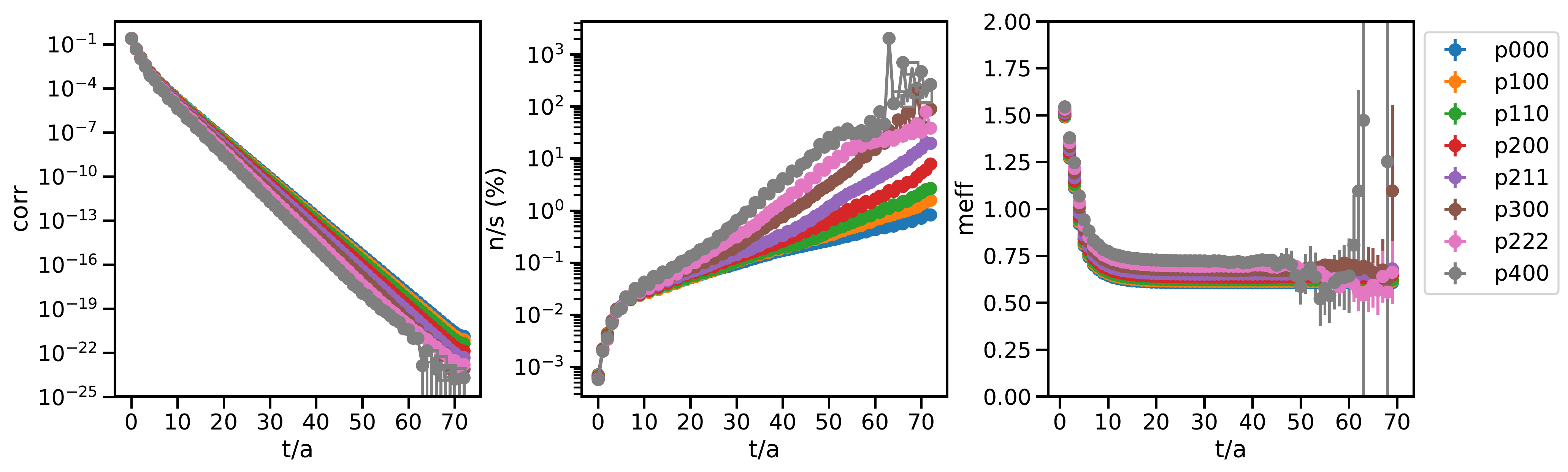}
\caption{Two-point functions used in the calculation of $B_s \to D_s \, \ell \, \nu$,
shown here for the 0.06 fm-0.1 $m_s$ ensemble.
The top panel shows, moving from left to right, the correlation functions, noise-to-signal, and effective masses for $H_s$ correlation functions, for a range of heavy input masses $m_h$. The bottom panels display the same information for two-point functions of the final state $D_s$ particle, over a range of momenta.}
\label{fig:2pt}
\end{figure}

In Fig.~\ref{fig:2pt} we show example two-point correlators from our analysis, computed here on the $a \approx 0.06$ fm $m_l/m_s=0.1$ ensemble.
For the heavy $H_s$ hadrons, we compute two-point correlators for a range
of heavy input masses: $m_h \approx 0.9 m_c, 1.0 m_c, 2.0 m_c, 3.0 m_c, 4.0 m_c$. These are shown in the top panel of the figure. From the second panel one can see that the noise-to-signal increases as the heavy mass increases, but overall the statistical precision is very good and long stable plateaus are observed in the effective masses in the third panel. The lower panel shows the final-state hadron (in this case $D_s$) for all of the different momenta studied. As expected the noise-to-signal increases with increasing recoil momentum, but stable effective masses are obtained even at our highest momentum studied.

\begin{figure}
    \centering
    \includegraphics[width=0.8\textwidth]{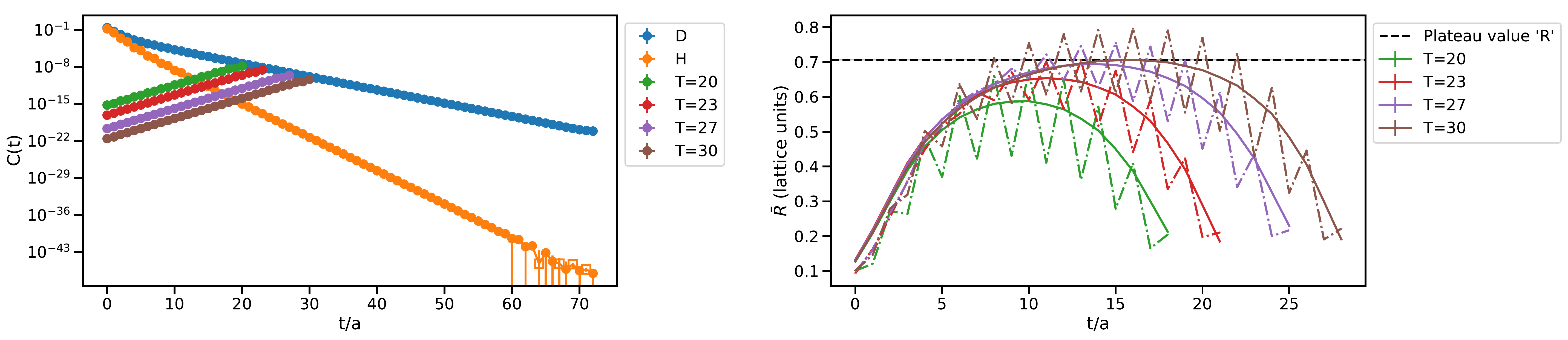}
    \caption{(Left) Combination of two- and three-point functions which are fit simultaneously to determine matrix element~\eqref{eq:Smxel}. (Right) Correlator data re-expressed in a convenient ratio form that visually illustrates the value extracted for the matrix element as a plateau (dotted line). Note that in practice the matrix elements are determined from a simultaneous multi-exponential fit of the correlators, as described in the text.}
    \label{fig:3pt}
\end{figure}

Fig.~\ref{fig:3pt} plots an example set of two-point and three-point functions, which are fit simultaneously according to 
Eq.~\eqref{eq:3pt_spectral_decomp} 
to extract matrix elements used in determining the form factors.
The right-hand plot shows the ratio of Eq.~\eqref{eq:ratio} from these correlators.

\subsection{Form factors}
Having extracted scalar and vector three-point matrix elements from simultaneous fits, we can relate these to the decay form factors via  
\begin{align}
\bra{L} \mathcal{V}^\mu \ket{H}
	&\equiv \sqrt{2 M_H} \left( v_H^\mu f_\parallel(q^2) + p_\perp^\mu f_\perp(q^2) \right)\\
	&\equiv f_+(q^2) \left( p_H^\mu + p_L^\mu - \frac{M_H^2 - M_L^2}{q^2} q^\mu \right) + f_0(q^2) \frac{M_H^2 - M_L^2}{q^2}q^\mu\\
\bra{L} \mathcal{S} \ket{H}
	&= \frac{M_L^2 - M_H^2}{m_h - m_\ell} f_0(q^2)	\,.
 \label{eq:Smxel}
\end{align}
In these expressions, $M_H$, $M_L$, $p_H^\mu$, and $p_L^\mu$ refer to the mass and four-momentum of the heavy initial (H) and light final state (L) mesons; $m_h$ and $m_\ell$ refer to the heavy and light input quark masses of the transition current; $v_H^\mu = p_H^\mu / M_H$ is the four-velocity of the heavy meson; $p_\perp^\mu = p_L^\mu - (p_L \cdot v_H) v_H^\mu$ is the component of the light hadron's momentum orthogonal to $v_H$, and; $q^\mu = p_H^\mu - p_L^\mu$ is the momentum transfer.
The final equality relating $f_0$ to the scalar matrix element follows from partial conservation of the vector current (see Eq.~(\ref{eq:PCVC}) below).
The manifestly covariant expressions simplify in the rest frame of the decaying heavy meson and take the following simple forms:
\begin{align}
f_\parallel	&= Z_{V^0}\frac{\matrixel{L}{V^0}{H}}{\sqrt{2 M_H}} \label{eq:f_parallel}\\
f_\perp		&= Z_{V^i}\frac{\matrixel{L}{V^i}{H}}{\sqrt{2 M_H}} \frac{1}{p^i_L} \label{eq:f_perp}\\
f_0				&= Z_{S}\frac{m_h-m_\ell}{M_H^2 - M_L^2} \matrixel{L}{S}{H}. \label{eq:f_0}
\end{align}

\begin{figure}[t]
    \centering
    \includegraphics[width=0.63\textwidth]{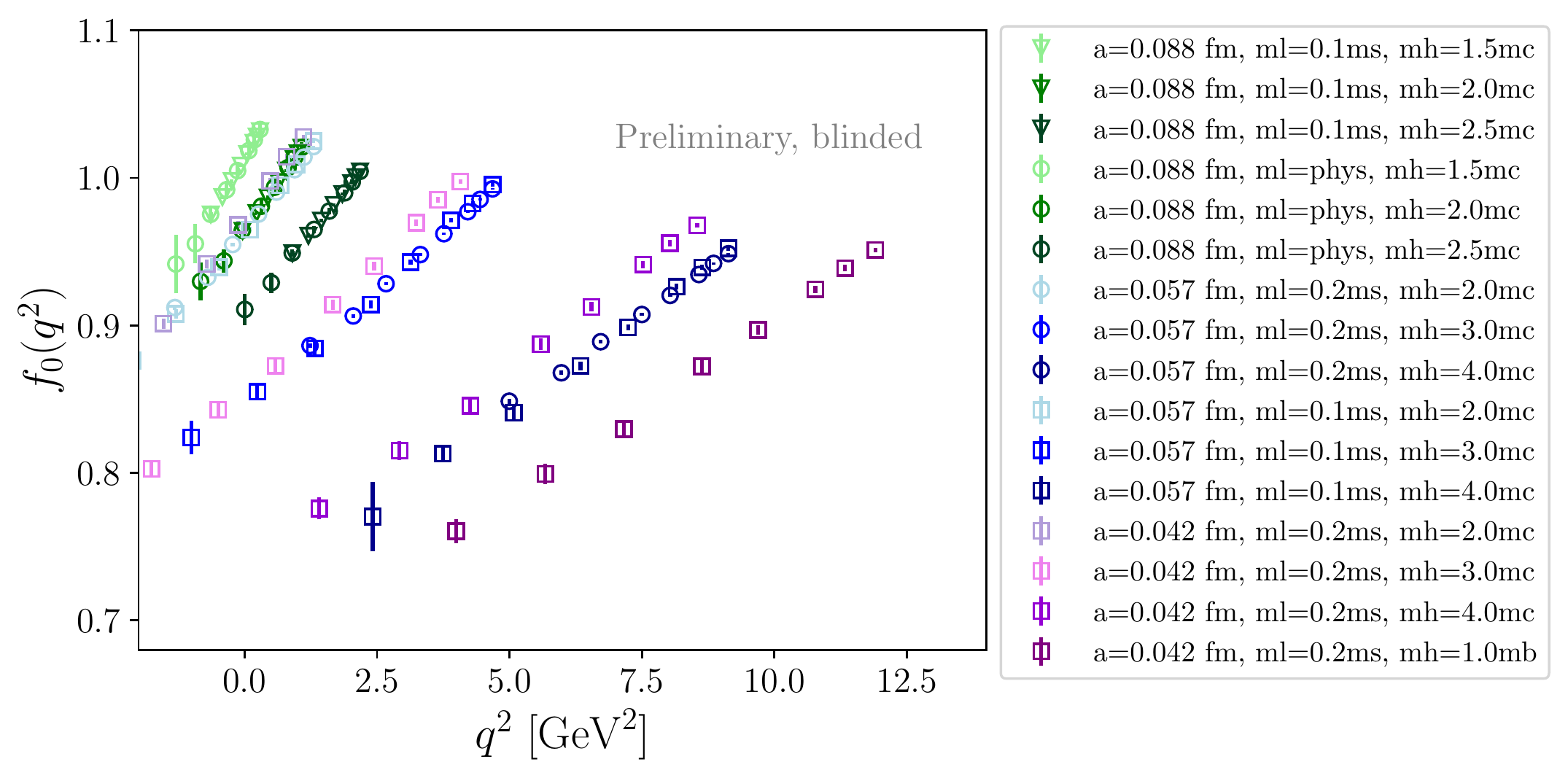}
    \caption{$f_0$ form factor for $H_s \to D_s$ decay as a function of momentum transfer $q^2$, for a range of heavy input masses.}
    \label{fig:f0}
\end{figure}

In Fig.~\ref{fig:f0}, we show results for the $f_0$ form factor of
$H_s \to D_s$ decay, plotted as a function of momentum transfer $q^2$.
In this and subsequent figures, the colors (green, blue, purple) correspond to the lattice spacings ($a \approx 0.09, 0.06, 0.042$ fm), and lighter/darker shading correspond to lighter/heavier input masses $m_h$.
As we go to finer lattice spacings, larger values of $m_h$ are accessed and the total energy available for the recoiling hadron increases. We observe small statistical errors over the kinematic range studied. The different symbol shapes in the figure correspond to different light-quark sea masses, and one observes very little sea-quark mass dependence in this decay.

\begin{figure}[t]
    \centering
    \includegraphics[width=0.85\textwidth]{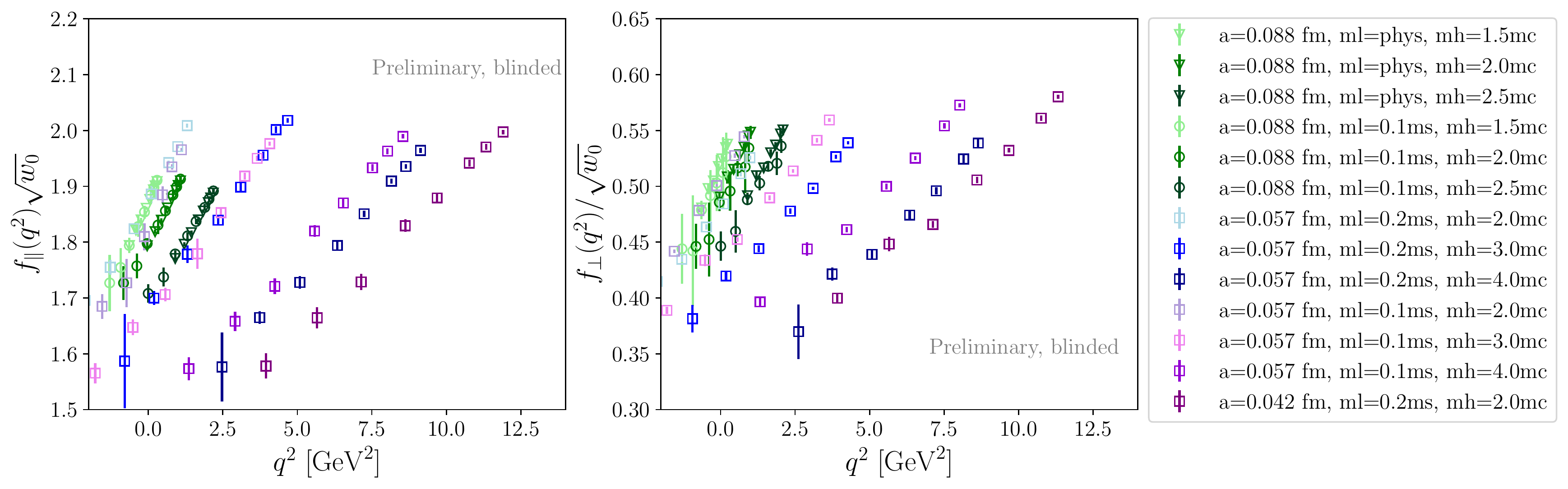}
    \caption{$f_{\parallel}$ (left) and $f_{\perp}$ (right) form factors for $H_s \to D_s$ decay
    as a function of momentum transfer $q^2$, for a range of heavy input masses.}
    \label{fig:fpar-fperp}
\end{figure}

In Fig.~\ref{fig:fpar-fperp} we show the $f_\parallel$ and $f_\perp$ form factors extracted using three-point functions involving the local-temporal and one-link spatial currents, respectively. Qualitatively these share similar features with the $f_0$ results, we observed good statistical control out to large momenta, and there is little evidence of light sea-quark mass effects.
The data plotted here is shown before renormalization of the vector current.
In order to take the chiral-continuum limit of this data the currents must be renormalized and we discuss this further in the next subsection.

\begin{figure}[h]
    \centering
    \includegraphics[width=0.65\textwidth]{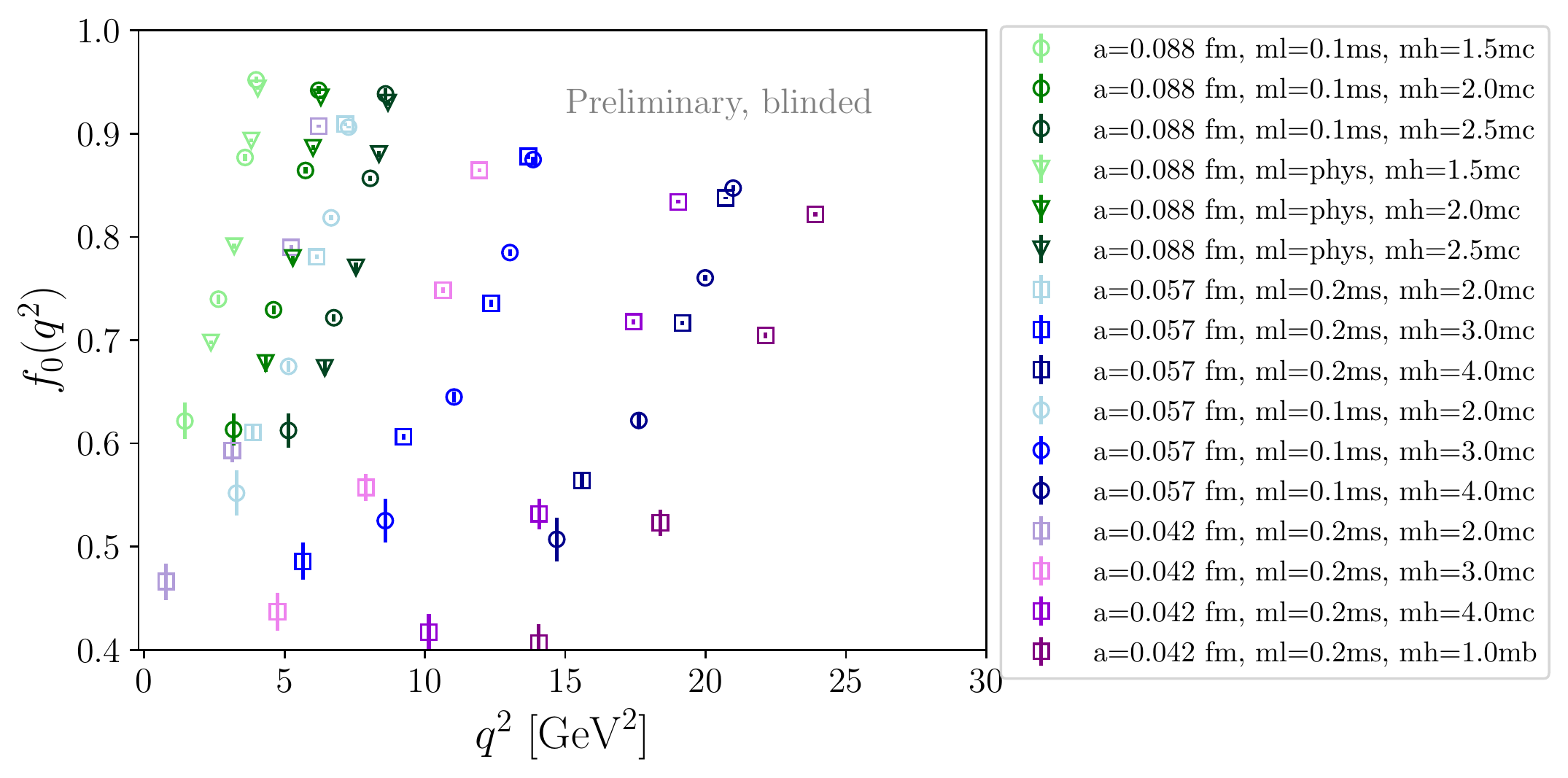}
    \caption{$f_0$ form factor for $H_s \to K$ decay as a function of momentum transfer $q^2$, for a range of heavy input masses.}
    \label{fig:BstoK-f0}
\end{figure}

Fig.~\ref{fig:BstoK-f0} shows $f_0(q^2)$ for the decay $H_s \to K$.
Here again we see good statistical control through most of the kinematic range. The light-quark mass dependence again appears to be relatively small, although here we expect some impact from the light valence quark in the kaon, which is matched to the light sea-quark mass.

\subsection{Vector current renormalization}
\begin{figure}[]
    \centering
     \includegraphics[width=0.65\textwidth]{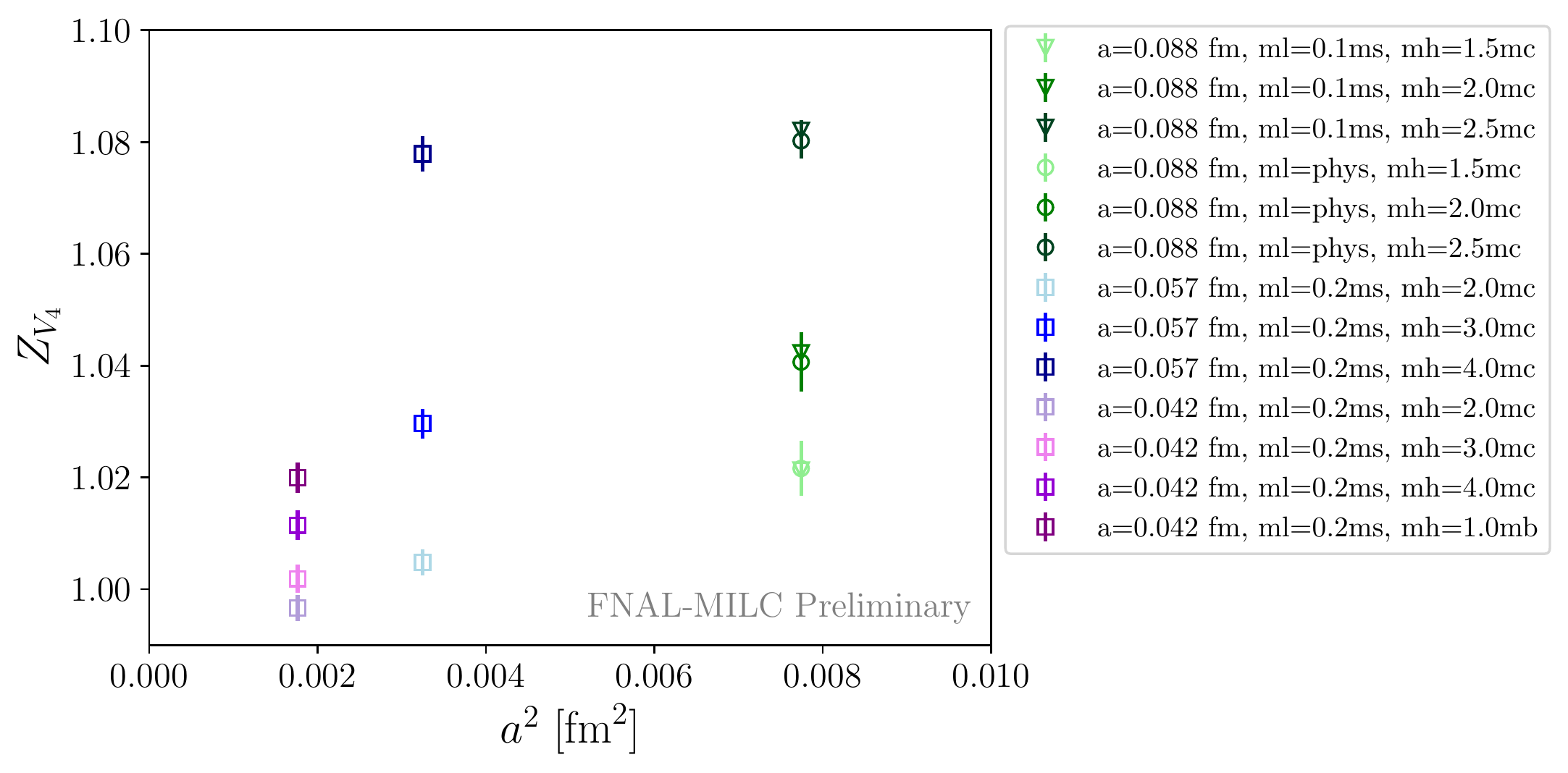}
    \caption{Renormalization factors for the local temporal vector current determined from Ward identities of three-point correlators.}
    \label{fig:Zs}
\end{figure}

The vector current operators we use to induce the quark transition require renormalization.
We renormalize the vector current by applying the partially
conserved vector current (PCVC) relation directly to extracted matrix elements:
\begin{equation}
\partial_\mu V^{\text{cons}}_{\mu} = (m_h - m_l) S \,.
\end{equation}
Applied to our lattice matrix elements,
\begin{align}
    Z_{V^4}(M_H - E_L) \bra{L} V^0 \ket{H}
    + Z_{V^i} \mathbf{q}\cdot \bra{L}\mathbf{V}\ket{H}
    = (m_h - m_l) \bra{L} S \ket{H} \label{eq:PCVC}\,,
\end{align}
where $V^4$ is local and $V^i$ is a one-link current.
At zero recoil momentum, the second term in~\eqref{eq:PCVC} is
zero and so $Z_{V^4}$ may be straightforwardly determined from the zero-recoil three-point matrix elements. 
In Fig.~\ref{fig:Zs} we show the results of this for the $B_s \to D_s$ correlators. Note that the same renormalization factors can also be used for $B \to D$ since the $h \to c$ quark transition is the same.
To find $Z_{V^i}$ non-zero recoil three-point matrix elements must be used in Eq.~\ref{eq:PCVC}. 
As a preliminary step we used
the $\mathbf{n}=(3,0,0)$ matrix elements along with the values for $Z_{V^4}$ computed in the previous step,
and found qualitative behavior similar to what is observed
for $Z_{V^4}$,
namely that factors tend towards 1 as the continuum is
approached, and also increase with increasing heavy quark mass
$am_h$, 
consistent with $(am_h)^n$-type lattice discretization errors.
A more robust determination, including fully quantified error bars,
may be obtained for $(Z_{V^4}, Z_{V^i})$ via simultaneous fit using all available momenta, and this is how we intend to obtain our final $Z$-factors. 

\section{Conclusion \& Outlook}
\vskip -0.25cm
We have presented preliminary results for semileptonic $B_{(s)}$-decay form factors, computed using the HISQ action for all valence quarks on the MILC collaboration's $N_f =2 +1 +1 $ HISQ ensembles. This is
an update to last year's proceeding~\cite{FermilabLattice:2021bxu}, where decays of $D$ mesons were also considered.  
The analysis of $D$-meson decay form factors was recently completed~\cite{FermilabLattice:2022gku}.
The calculations outlined here together with high-precision experimental measurements will result in improved determinations of $|\Vcs|$, $\Vcb$, $|\Vcd|$, $\Vub$, providing some of the most stringent tests of the Standard Model in these respective sectors.

To access $B$ physics in the continuum we work at a range of heavy
mass input values on each ensemble, with $am_h \lesssim 1$.
On the finest ensembles studied ($a=0.042$ and 0.03 fm) we have simulated valence quarks at or near to the physical $b$-quark mass.
Treating all quarks with the same action allows us to compute and analyze data for a range of decays in a unified treatment, with non-perturbatively renormalized currents. For the decays considered we observe good statistical control over the kinematic range studied. We have simulated on ensembles with
physical light sea-quarks at our $a=0.09$ and 0.06 fm lattice spacings,
which should allow chiral interpolation, thereby reducing systematic errors.
In the future we plan extend this to the physical mass ensemble at $a=0.042$ fm. We are currently analyzing the data computed at our finest lattice spacing of $a=0.03$ fm.

\vskip -0.25cm
\section{Acknowledgments}
\vskip -0.25cm
This document was prepared by the Fermilab Lattice and MILC Collaborations using the resources of the Fermi National Accelerator Laboratory (Fermilab), a U.S. Department of Energy, Office of Science, HEP User Facility.
Fermilab is managed by Fermi Research Alliance, LLC (FRA), acting under Contract No. DE-AC02-07CH11359.
This material is based upon work supported
by the U.S. Department of Energy, Office of Science under grant Contract Numbers DE-SC0015655 (A.L., A.X.K.), DE-SC0010120 (S.G.), DE-SC0011090 (W.J.), and DE-SC0021006 (W.J.); 
by the Simons Foundation under their Simons Fellows in Theoretical Physics program (A.X.K.);
by the U.S. National Science Foundation under Grants No.\ PHY17-19626 and PHY20-13064 (C.D., A.V.);
by SRA (Spain) under Grant No.\ PID2019-106087GB-C21 / 10.13039/501100011033 (E.G.);
by the Junta de Andalucía (Spain) under Grants No.\ FQM-101, A-FQM-467-UGR18 (FEDER), and P18-FR-4314 (E.G.);
and by AEI (Spain) under Grant No.\ RYC2020-030244-I / AEI / 10.13039/501100011033 (A.V.).

Computations for this work were carried out in part on facilities of the USQCD Collaboration, which are funded by the Office of Science of the U.S. Department of Energy.
An award of computer time was provided by the Innovative and Novel Computational Impact on Theory and Experiment (INCITE) program. This research used resources of the Argonne Leadership Computing Facility, which is a DOE Office of Science User Facility supported under contract DE-AC02-06CH11357. This research also used resources of the Oak Ridge Leadership Computing Facility, which is a DOE Office of Science User Facility supported under Contract DE-AC05-00OR22725.
The authors acknowledge support from the ASCR Leadership Computing Challenge (ALCC) in the form of time on the computers Summit and Theta.
This research used resources of the National Energy Research Scientific Computing Center (NERSC), a U.S. Department of Energy Office of Science User Facility located at Lawrence Berkeley National Laboratory, operated under Contract No. DE-AC02-05CH11231.
This work used the Extreme Science and Engineering Discovery Environment (XSEDE), which is supported by National Science Foundation grant number ACI-1548562.
This work used XSEDE Ranch through the allocation TG-MCA93S002~\cite{XSEDE}.
The authors acknowledge the \href{http://www.tacc.utexas.edu}{Texas Advanced Computing Center (TACC)} at The University of Texas at Austin for providing HPC resources that have contributed to the research results reported within this paper.
This research is part of the Frontera computing project at the Texas Advanced Computing Center. Frontera is made possible by National Science Foundation award OAC-1818253~\cite{Frontera}.


\end{document}